\documentclass{PoS}
\usepackage[outercaption]{sidecap}
\title{Forward photon measurements in ALICE as a probe for low-x gluons}

\ShortTitle{Forward photon measurements in ALICE}

\author{\speaker{Thomas Peitzmann}, on behalf of the ALICE Collaboration\\
        Utrecht University/Nikhef, Princetonplein 1, 3584CC Utrecht, The Netherlands\\
        E-mail: \email{t.peitzmann@uu.nl}}
\abstract{The low-$x$ gluon density in the proton and, in particular, in nuclei is only very poorly constrained, while a better understanding of the low-$x$ structure is crucial for measurements at the LHC and also for the planning of experiments at future hadron colliders. In addition, deviations from linear QCD evolution are expected to appear at low $x$, potentially leading to gluon saturation and a universal state of hadronic matter, the color-glass condensate. However, these effects have not been unambiguously proven to date. Fortunately, data from the LHC can be used directly to provide better constraints of the parton distribution functions (PDFs). In this context, a Forward Calorimeter (FoCal) is proposed as an addition to the ALICE experiment, to be installed in the Long Shutdown 3.
  
The main goal of the FoCal proposal is to measure forward direct photons in pp and p--Pb collisions to obtain experimental constraints on proton and nuclear PDFs in a new region of low $x$. Based on the current knowledge from DIS experiments and first results from LHC, we will discuss the physics case for this proposed detector. While open charm measurements do provide important constraints, a photon measurement would provide additional unique information.  The direct photon measurement requires a new electromagnetic calorimeter with extremely high granularity. The corresponding innovative design principle of a high-resolution Si-W sandwich calorimeter is discussed. }

\FullConference{International Conference on Hard and Electromagnetic Probes of High-Energy Nuclear Collisions\\
		30 September - 5 October 2018\\
		Aix-Les-Bains, Savoie, France}

\begin{document}

\section{Parton distributions at small $x$}
\begin{figure}[htb]
\begin{center}
\includegraphics[width=0.8\textwidth]{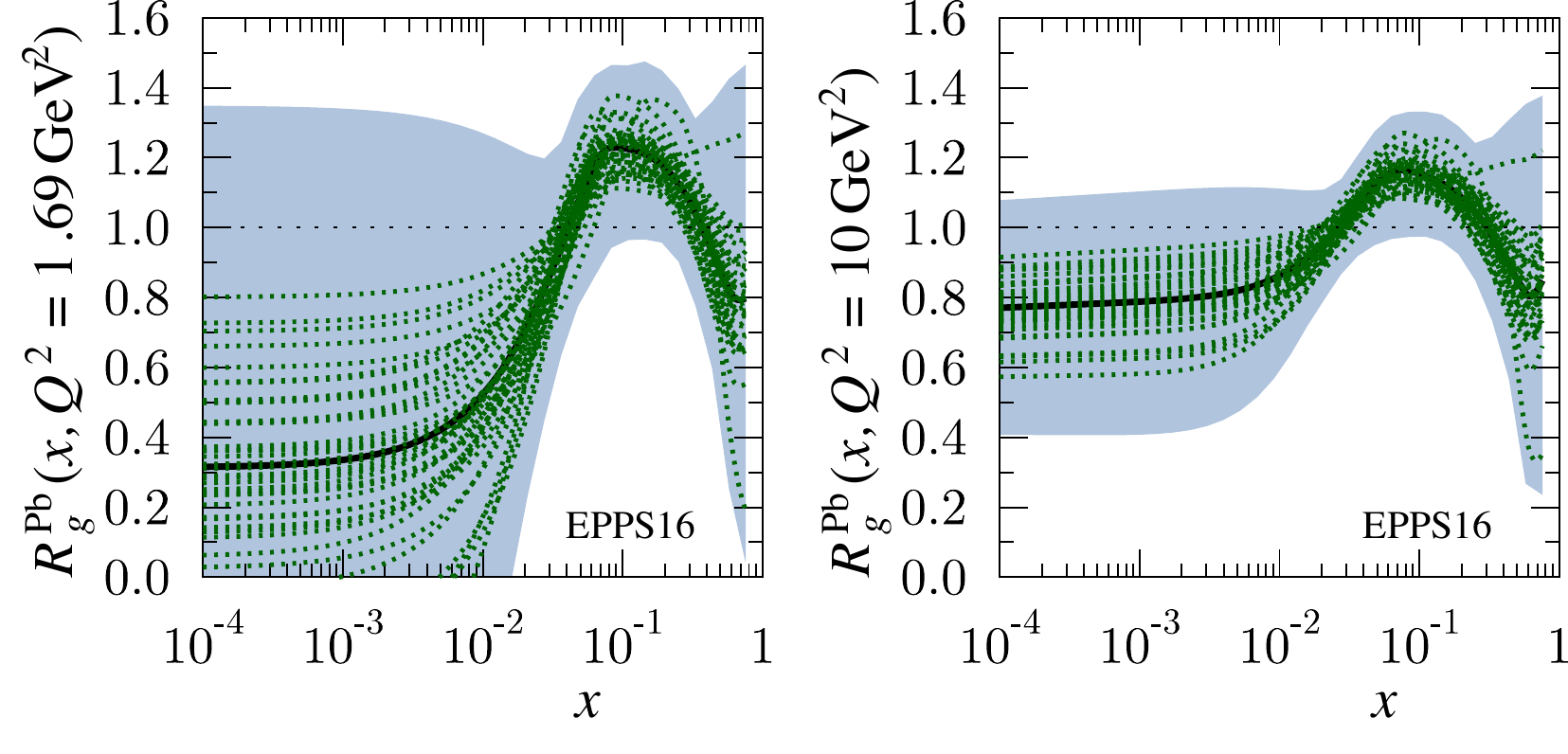}
\caption{\protect\label{fig:epps16}  Uncertainties of nuclear modification factors of gluon PDFs from EPPS16 \cite{epps16}.} 
\end{center}
\end{figure}

The internal structure of hadrons at low momentum fraction $x$ is a research topic with many open questions that has received a lot of attention in the recent past. While the parton densities at low $x$ in hadrons and nuclei are crucial for the quantitative description of hadron collisions at high energy, 
they are only very poorly known.  At moderate $Q^2$, there are no direct experimental constraints on the PDFs for protons at $x < 10^{-4}$ and for nuclei at $x < 10^{-2}$. Shadowing of nuclear PDFs has been observed, but it is unknown how this develops towards smaller values of $x$ --- this is particularly severe, because it contributes to an uncertainty in the interpretation of the nuclear modification of particle yields observed in nucleus-nucleus collisions. Figure~\ref{fig:epps16} illustrates the uncertainty on the gluon nuclear PDFs. In particular for low $Q^2$ (left), the nuclear modification factor $R_g$ is essentially unknown. This improves for intermediate $Q^2$ (right), still the uncertainty is considerable. One should also note, that the example functional shapes shown as green dashed lines in the figure share a common feature: there is almost no $x$-dependence of $R_g$ for $x < 10^{-2}$. This limits the sensitivity of PDF fits for very low $x$, and possibly obscures the true influence of a given measurement on further constraining the PDFs, as we will discuss below.  

\vspace{-5mm}
Linear QCD evolution (DGLAP) is used to extrapolate to the inaccessible kinematic regions. However, the increase of parton densities towards low $x$ cannot continue indefinitely. Non-linear effects are expected to ``tame'' the gluon density and lead to \textit{gluon saturation} --- these effects are enhanced in nuclei compared to protons because of the higher parton density. Models of gluon saturation, like the \textit{colour glass condensate}, provide alternative quantitative descriptions of the initial state of hadrons and nuclei at very high energy \cite{CGC}. Gluon saturation is consistent with many phenomena observed in high-energy scattering, like shadowing, however, no unambiguous proof has been obtained so far. DIS is currently limited in its kinematic reach to provide such a proof, and for many of the existing observables, alternative explanations are possible. 

\begin{figure}[htb]
\begin{center}
\includegraphics[width=0.35\textwidth]{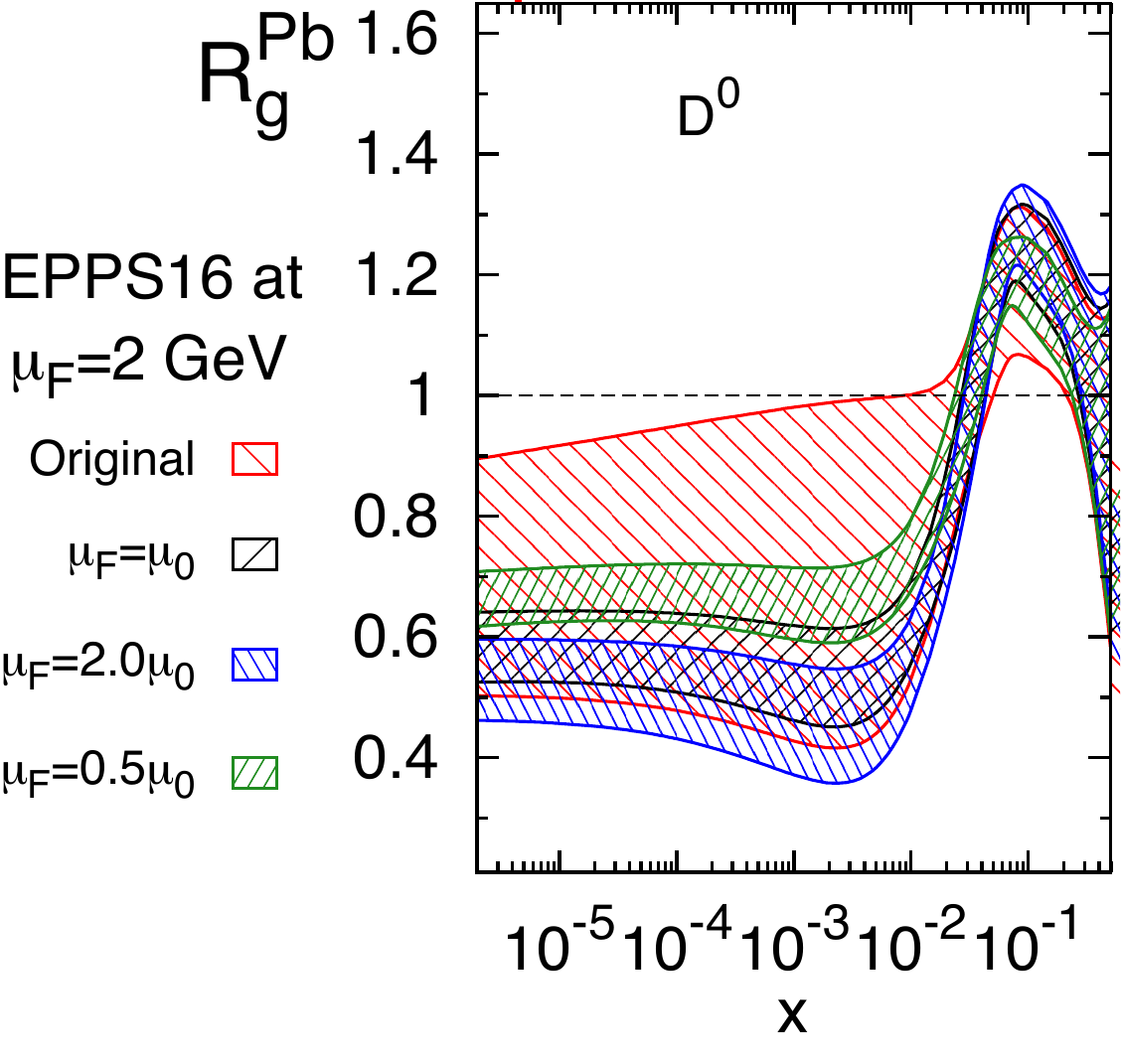}\hfill%
\includegraphics[width=0.50\textwidth]{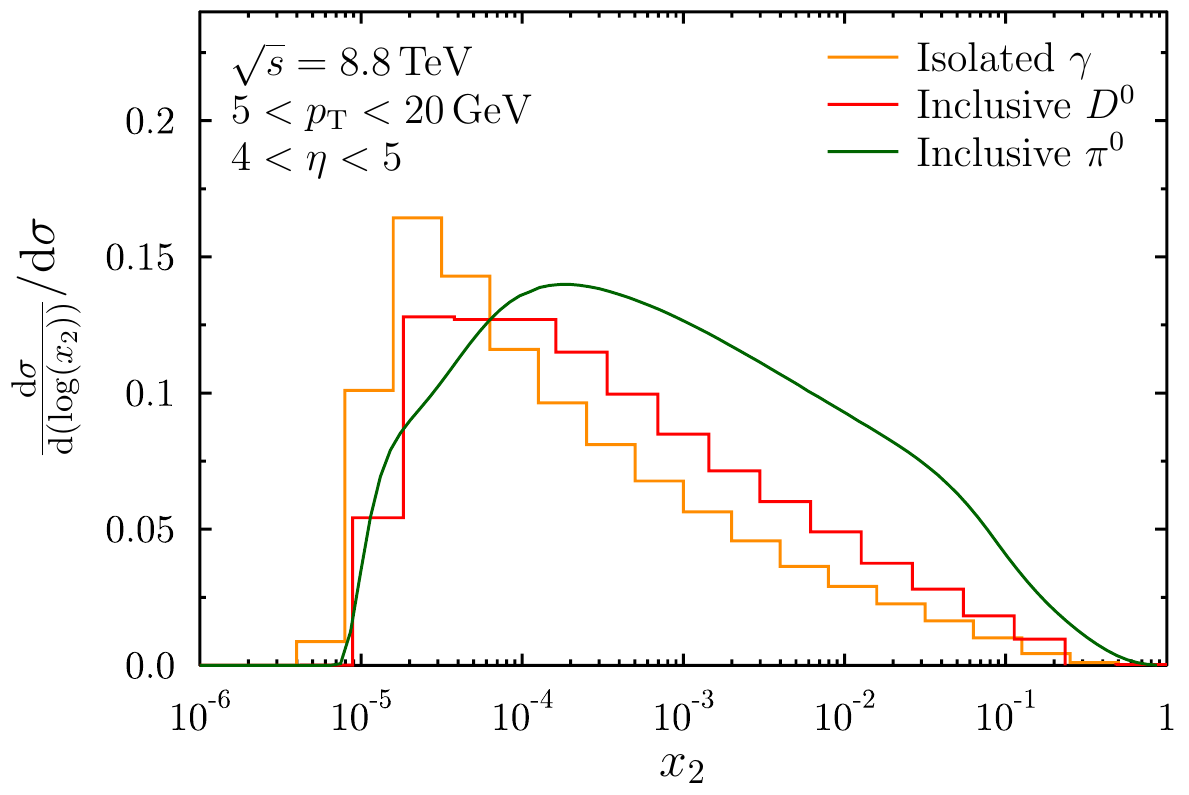}
\caption{\protect\label{fig:charm-compare} Left: Nuclear modification factor of the EPPS16 gluon PDF without (red hashed area) and with (black, blue, and green) constraints from forward D$^0$ measurements by LHCb as given in \cite{kusina} . 
Right: Distributions of $x_2$ for the production of different particles ($\pi^0$, D$^0$, and $\gamma$) for the same kinematic conditions at forward rapidity from NLO pQCD calculations \cite{ilkka}.} 
\end{center}
\end{figure}

Recently, the possibility has been realised of using data from hadron colliders to constrain the PDFs, or learn about effects of gluon saturation.  
In particle production in hadronic reactions, the minimally reachable $x$ values can be estimated from leading-order kinematics as $x \approx 2 p_\mathrm{T} e^{-y} / \sqrt{s}$, so in particular forward particle production at the highest beam energy, i.e.\ at the LHC, is of interest.
Most measurements have used hadronic observables, which bring additional uncertainties. In particular for light hadrons, the elementary production processes are not very well under control, and fragmentation leads to a bias towards larger values of $x$. The most promising case in the hadronic sector is the production of heavy-flavour hadrons, which has been employed in recent PDF studies \cite{kusina}. The strongest constraints come from the measurement of forward D$^0$ production by the LHCb Collaboration \cite{lhcb-d}, where they observe a significant suppression in p--Pb compared to pp. The consequence of including such data for PDF constraints can be seen in the left panel of Fig.~\ref{fig:charm-compare}, which shows the uncertainties of the nuclear modification $R_{\mathrm{g}}^{\mathrm{Pb}}$ of the gluon density. While the uncertainties are large before using the LHC heavy-flavour data, they shrink considerably when these data are included, and the central value moves to the lower limit of the previous uncertainty band. The values do show a significant scale dependence. Another important feature of these descriptions of $R_{\mathrm{g}}^{\mathrm{Pb}}$ is the lack of $x$ dependence for low values, similar to what was observed in Fig.~\ref{fig:epps16}, most likely caused by the choice of the analytical shapes of the parameterisations used \cite{epps16}. 
%It is likely, that these additional constraints on the PDFs do not make specific use of information related to the very low values of $x$. 
With the given shapes, constraints at rather moderate values of $x$ ($\approx 10^{-3}$) would by construction limit these functions also at much lower values.
\vspace{-1.mm}

\vspace{-5mm}
In fact, the measurements in question are sensitive to a broad range of $x$ values. Figure~\ref{fig:charm-compare} shows on the right the $x$ distributions probed by different forward measurements in collisions at the LHC as calculated by NLO pQCD \cite{ilkka}. All distributions share a similar low edge, which is close to the leading-order estimate cited above, but the distributions have a significant tail towards larger values. Neutral pions have the strongest contribution at large $x$, and D$^0$ mesons are more concentrated towards somewhat lower $x$, but the best low-$x$ probe among these three would be provided by direct isolated photons.
In addition, direct photons have the advantage, that the production processes are well understood, and that they do not suffer any final state modification, unlike light and possibly also heavy hadrons.

\section{Forward direct photon measurements}

\begin{figure}[htb]
\begin{center}
\includegraphics[width=0.9\textwidth]{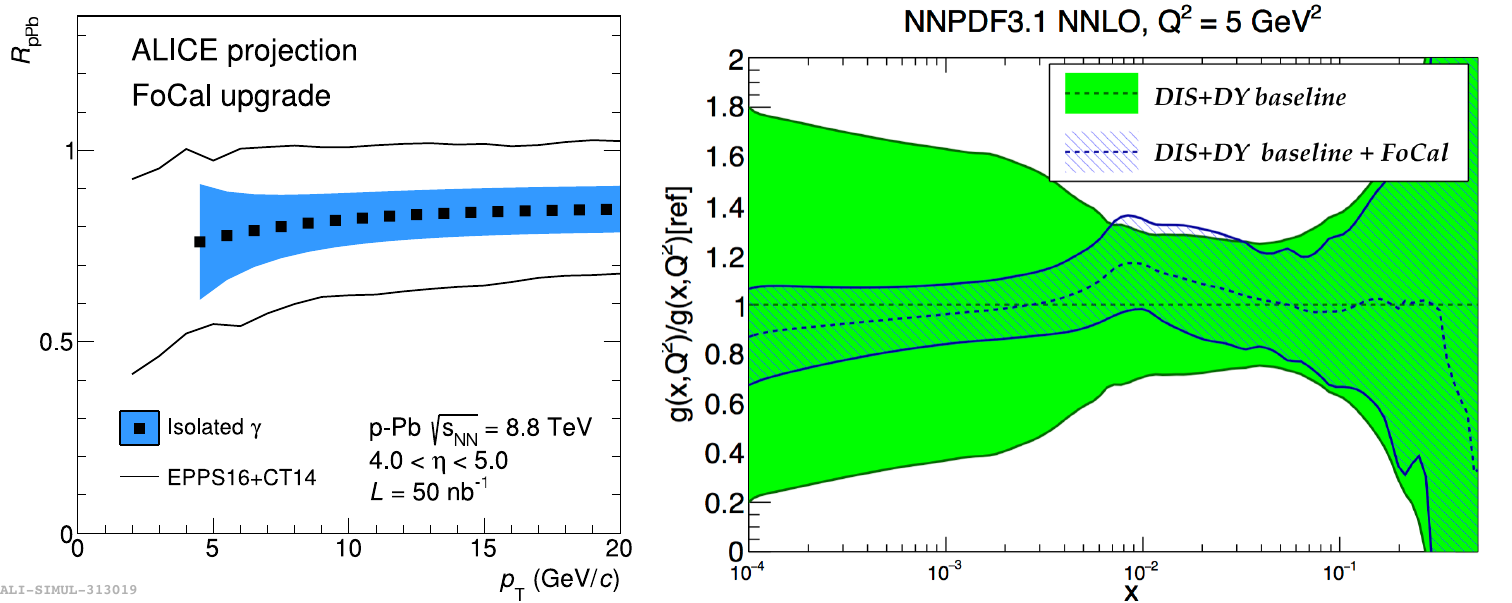}
\caption{\protect\label{fig:nucl-mod}  Left:  Nuclear modification factor $R_{\mathrm{pPb}}$ of isolated photons to be measured with FoCal (blue band). The black curves show the current EPPS16 PDF uncertainty, and the blue band the estimated experimental uncertainty. Right:  Estimate of the uncertainties on gluon nuclear PDFs from two scenarios: a fit to existing nuclear DIS and DY (green band) and a fit to DIS and DY including pseudo-data from a FoCal measurement (blue hashed band) \cite{rojo}.} 
\end{center}
\end{figure}

We propose to measure isolated direct photons at forward rapidity in pp and p--Pb collisions within the ALICE experiment at the LHC. Detailed GEANT simulations using a forward calorimeter (FoCal) as described in the next section have been performed. The main result on the detector performance is displayed in the left panel of Fig.~\ref{fig:nucl-mod}, which shows an estimate of the systematic uncertainty (blue band) of a measurement of the nuclear modification factor $R_{\mathrm{pPb}}$ of isolated photons. The statistical uncertainties are smaller than the symbols. The black solid lines indicate the current uncertainty estimate for NLO pQCD calculations using the EPPS16 nuclear PDFs, which are significantly larger than those  expected from the proposed measurement.

\vspace{-5mm}
The impact of such a measurement has also been investigated in the NNPDF framework \cite{rojo}. The right panel of Fig.~\ref{fig:nucl-mod} shows uncertainty estimates for two scenarios relevant for the nuclear PDFs. The green band shows how far the gluon density is constrained by DIS and Drell-Yan data as they are available for nuclei. For the hashed blue band constraints from pseudo-data of the FoCal detector, as derived from the performance simulations, have been used in addition. No specific parameterisation of a nuclear modification factor has been used. Clearly, an isolated photon measurement would lead to a significant reduction of the uncertainties, in particular at low $x$.

Such measurements of direct photons at $3.5 < y < 5$ for $p_{\mathrm{T}} \approx 4-20 \, \mathrm{GeV}/c$ will not be possible with existing detectors. In particular the discrimination of direct from decay photons, which needs spatial resolving power for the $\pi^0$ decays, requires a calorimeter of very small Moli\`ere radius and extremely fine granularity, beyond state-of-the-art technology.

\begin{SCfigure}
\includegraphics[width=0.5\textwidth]{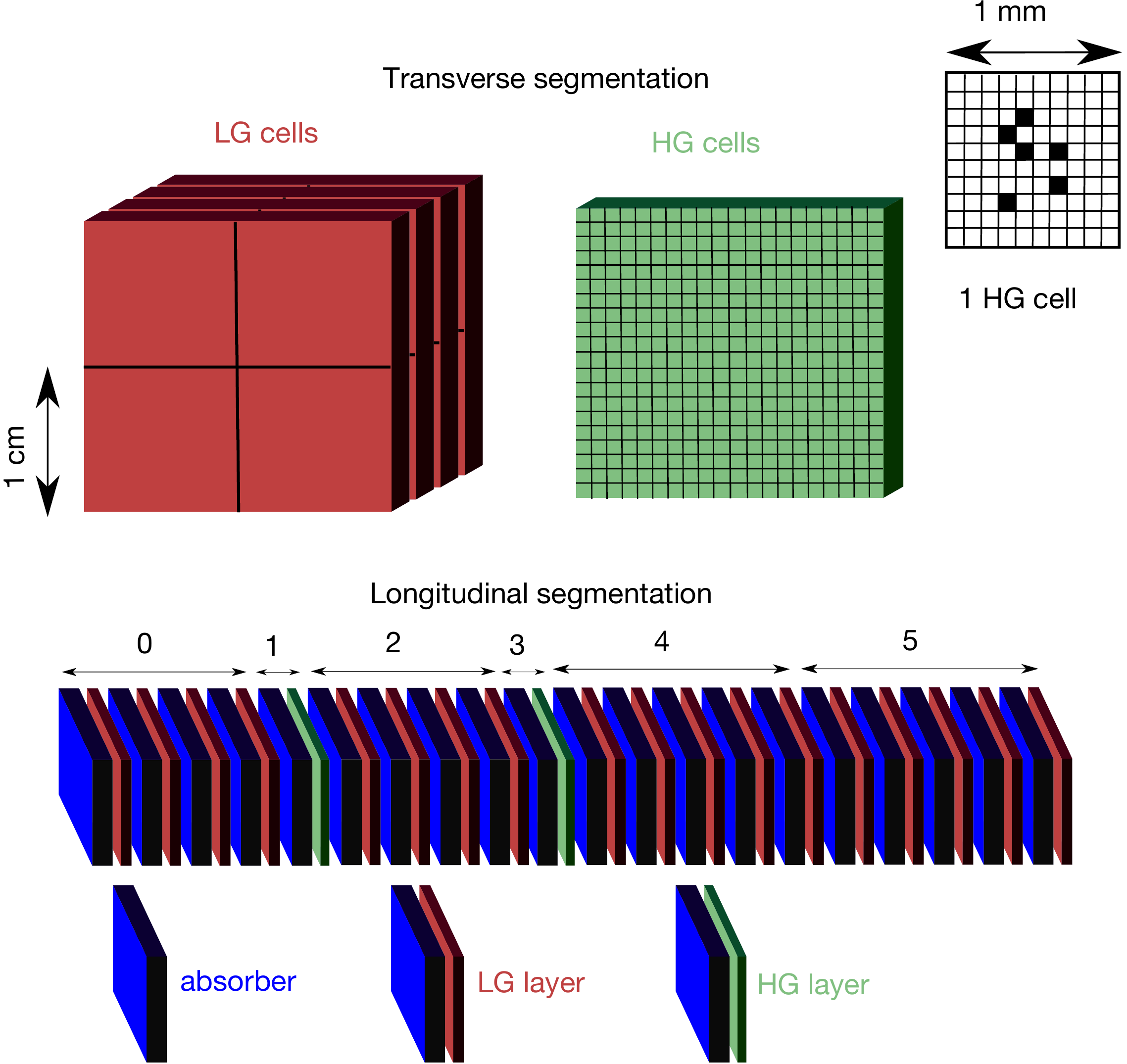}
\caption{Left: Schematic view of the internal structure of the FoCal detector with W-absorbers and Si-sensor layers of low (LG) and high (HG) granularity.
   \label{fig:detector}}
\end{SCfigure}

Within the ALICE Collaboration, it has been proposed to install a forward calorimeter (FoCal) at  $\approx 7$~m distance from the interaction point as an upgrade in the LHC Long Shutdown 3. The detector, covering $\approx 1$~m$^2$, would consist of a Si-W sandwich structure with a number of layers of extremely high granularity (HG), which would provide the two-shower separation capability, and layers of somewhat lower granularity (LG) more suited for the total energy measurement. A sketch of the internal structure as currently implemented in simulations is shown in Fig.~\ref{fig:detector}. 

Extensive R\&D is being performed on both the LG and HG technologies. While the LG layers will likely use Si-pad sensors, the most promising candidate for the HG layers are CMOS sensors, which allow a high pixel density with little additional material, keeping the effective Moli\`ere radius small. 
%However, very little was known until recently about their behaviour in a calorimeter, where the typical pixel occupancies are much higher than in tracking applications. 
To explore this new technology, a prototype of a high-granularity electromagnetic calorimeter with a total number of 39 million pixels of $30 \times 30 \, \mu\mathrm{m}^2$ in a $4 \times 4 \times 11\,\mathrm{cm}^3$ volume and a Moli\`ere radius of $\approx 11$~mm has been built and tested successfully \cite{focal-jinst}. The results clearly demonstrate the extraordinary capabilities of two-shower separation. 
Also prototypes using pad sensors have been built and have been used in test beam measurements. The next R\&D steps are ongoing to develop a suitable readout for the pad sensors, and an improved pixel sensor based on the development for the ALICE Inner Tracking System upgrade \cite{alice-its}.

\vspace{-2mm}
In summary, the best opportunity in the near future to significantly constrain the low-$x$ structure of protons and nuclei and to shed light on the open question of gluon saturation is the measurement of direct photons at forward rapidity in p--A and pp at the LHC. While no existing detector would be able to perform this measurement, it should be possible with a proposed upgrade to the ALICE experiment with a Forward Calorimeter (FoCal). To allow this measurement, an extremely high granularity calorimeter beyond the state of the art has to be developed. Corresponding R\&D is ongoing, and the proof-of-principle measurements have been successfully performed.

\end{document}